\documentclass[conference]{IEEEtran}
\usepackage[pscoord]{eso-pic}
\IEEEoverridecommandlockouts
\usepackage{cite}
\usepackage[font=footnotesize,skip=3pt]{caption}
\usepackage{amsmath,amssymb,amsfonts}
\usepackage{algorithmic}
\usepackage{graphicx}
\usepackage{subcaption}
\usepackage{booktabs}
\usepackage{array}
\usepackage{multirow}
\usepackage{textcomp}
\usepackage{xcolor}
\usepackage{comment}
\usepackage{siunitx}
 \usepackage[font=footnotesize]{caption}
\usepackage[colorlinks=true, allcolors=blue]{hyperref}
\usepackage{orcidlink}
\usepackage[compact]{titlesec}
\titleformat{\paragraph}[runin]
  {\normalfont\normalsize\itshape}{}{0pt}{}
\titlespacing*{\section}{0pt}{6pt plus 2pt minus 2pt}{3pt}
\titlespacing*{\subsection}{0pt}{5pt plus 2pt minus 1pt}{2pt}
\titlespacing*{\subsubsection}{0pt}{4pt plus 1pt minus 1pt}{2pt}
\titlespacing*{\paragraph}{0pt}{3pt plus 1pt minus 1pt}{0.5em}

\usepackage{enumitem}
\setlist{noitemsep, topsep=2pt, parsep=2pt, partopsep=0pt}

\def\BibTeX{{\rm B\kern-.05em{\sc i\kern-.025em b}\kern-.08em
    T\kern-.1667em\lower.7ex\hbox{E}\kern-.125emX}}

\begin{document}

\bstctlcite{BSTcontrol}

\title{Cross-Domain Generalization in Optical Networks via Joint Contrastive and Classification Learning\thanks{This work has been supported by the EUREKA CELTIC-NEXT SUSTAINET-Advance project, funded by the Swiss Innovation Agency Innosuisse No. 119.588 INT-ICT, and by Vinnova (Sweden's Innovation Agency) No. 2025-02987. Corresponding author: ali.alhousseini@supsi.ch \\ All source code is available on GitHub at the following link: \href{https://github.com/alialhousseini/QoT-transfer-learning}{https://github.com/alialhousseini/QoT-transfer-learning}}}
\author{
\IEEEauthorblockN{
Ali Al Housseini~\orcidlink{https://orcid.org/0009-0003-6682-474X}\IEEEauthorrefmark{1}\IEEEauthorrefmark{2},
Carlos Natalino~\orcidlink{0000-0001-7501-5547}\IEEEauthorrefmark{3},
Paolo Monti~\orcidlink{0000-0002-5636-9910}\IEEEauthorrefmark{3},
Omran Ayoub~\orcidlink{0000-0002-3884-3594}\IEEEauthorrefmark{1}
}
\IEEEauthorblockA{\IEEEauthorrefmark{1}
University of Applied Sciences and Arts of Southern Switzerland, Lugano, Switzerland}
\IEEEauthorblockA{\IEEEauthorrefmark{2}
University of Southern Switzerland, Lugano, Switzerland}
\IEEEauthorblockA{\IEEEauthorrefmark{3}
Department of Electrical Engineering, Chalmers University of Technology, Gothenburg, Sweden}

}

\maketitle

\begin{abstract}
The robustness of machine learning techniques across heterogeneous network domains remains an open challenge in optical networks. Models trained on data from a specific topology or operational configuration often exhibit degraded performance when deployed in unseen networks. In this work, we address this challenge by proposing a representation learning technique aimed at capturing task-relevant relationships that remain stable across domains. The proposed technique is based on a novel joint contrastive and classification learning approach in which representation learning and task optimization are performed simultaneously, allowing both objectives to shape the latent space. Experimental results on a representative use case, namely, lightpath quality of transmission estimation, demonstrate the effectiveness of our approach compared to baseline approaches, and highlight its capacity for rapid adaptation, providing excellent performance even with limited fine-tuning.
\end{abstract}


\section{Introduction}

Machine learning (ML) techniques are increasingly explored for supporting data-driven decision-making in optical networks, with promising results in controlled settings \cite{Gao_2024_GeneralizationCognitiveOptical, Sadighi_2025_GeneralizabilityMLBasedClassification, Liu_2025_ModuleEnhanceGeneralization}. However, models trained on a specific network topology, operational configuration, or lightpath often exhibit degraded performance under unseen conditions, revealing a persistent generalization challenge across heterogeneous domains \cite{usmani2022transfer, usmani2024integrating, zhou2023neuron, zhou2024evolutionary, Lechowicz_2025_QoTEstimationMarginDriven, aladin2025automated}.

This lack of generalizability, while being a common issue in ML across all domains, is particularly pronounced in optical networks due to the strong dependence of data distributions on network-specific characteristics.
Indeed, optical network data can be shaped by multiple factors, including network topology, link length distributions, physical-layer impairments, manufacturing differences in devices, and operator- or vendor-specific configurations.
For instance, each amplifier has a different power profile which impacts the accuracy of ML models \cite{Wang_2024_MultiSpanOpticalPower}.
Moreover, research has shown that the frequency/wavelength of a channel changes the response to state-of-polarization sensing \cite{Sadighi_2025_GeneralizabilityMLBasedClassification}.
As a consequence, domain shifts between training and operation environments are common and unavoidable. At the same time, collecting large volumes of labeled data for each new network is costly and often impractical, limiting the feasibility of fully training completely new models for each network condition.

\newcommand{\placetextbox}[3]{
  \setbox0=\hbox{#3}
  \AddToShipoutPictureFG*{
    \put(\LenToUnit{#1\paperwidth},\LenToUnit{#2\paperheight})%
    {\vtop{{\null}\makebox[0pt][c]{#3}}}}
  }
\placetextbox{.2}{0.055}{978-3-903176-78-2 \textcopyright\ 2026 IFIP}

Most existing ML approaches implicitly assume that training and test data are drawn from the same data distribution, an assumption that rarely holds in operational optical networks. Classical tabular ML models, such as tree-based ensembles, can achieve strong performance on a single network but lack explicit mechanisms to learn representations that are robust to domain shifts \cite{Gao_2024_GeneralizationCognitiveOptical, Sadighi_2025_GeneralizabilityMLBasedClassification}. Transfer learning has been proposed as a potential remedy; however, its effectiveness in this context is constrained. Pretraining on a source domain followed by fine-tuning on a target domain typically requires sufficient labeled target data and may suffer from overfitting or negative transfer when domain discrepancies are pronounced. 

In this work, we address the generalization challenge by proposing a \emph{representation learning} approach based on \emph{joint contrastive and classification learning}, aiming to capture relationships that remain stable across domains. The proposed approach jointly shapes the latent space, i.e., the internal feature representation where input data are mapped into a higher-dimensional task-relevant embeddings, through contrastive learning objectives while optimizing the classification task, encouraging representations that are compact within classes and well separated across classes. We hypothesize that the resulting representations align more closely with the underlying task structure, potentially improving robustness under domain shifts.

To evaluate our proposed approach, we consider lightpath Quality of Transmission (QoT) estimation as a representative use case, formulated as a binary classification problem. QoT estimation provides a concrete and practically relevant setting to study generalization across varying network conditions. Yet, the proposed methodology is not tied to this specific task and is applicable to a broader class of optical-network learning problems affected by similar generalization issues. Our experimental evaluation considers scenarios in which abundant data is available from a source network, while no or only limited labeled data is accessible from a target network. Through controlled target-domain data injection and comprehensive comparisons against state-of-the-art ML baselines, we demonstrate that the proposed methodology  achieves improved generalization performance. Specifically, our results highlight that the proposed joint optimization strategy successfully learns robust, task-relevant representations that remain stable across heterogeneous network domains.

\section{Related Work}
Generalization across heterogeneous network domains has emerged as a central challenge in the application of ML to optical networks.
\cite{Gao_2024_GeneralizationCognitiveOptical, Sadighi_2025_GeneralizabilityMLBasedClassification, Liu_2025_ModuleEnhanceGeneralization}.
Recent studies have investigated this challenge from complementary angles, including empirical quantification of domain shift, TL, architectural modifications, and representation learning.

Empirical investigations have systematically quantified the impact of domain shifts.
The work in \cite{Sadighi_2025_GeneralizabilityMLBasedClassification} shows that ML classifiers for State-of-Polarization event detection drop from 98.63\% to 8.11\% accuracy when transferred across spectral bands and fiber links, while \cite{Akbari_2025_LeveragingSharedData} reports that standard regression models for OSNR estimation fail to generalize in zero-shot scenarios across five multi-organization datasets.

TL has been widely explored to mitigate domain shift. Usmani et al. \cite{usmani2022transfer, usmani2024integrating} progressively refine TL for generalized signal-to-noise ratio (GSNR) estimation, from weight-based knowledge transfer across C-band configurations to knowledge-distillation-compressed student models with 93.6\% fewer trainable parameters. Zhou et al. \cite{zhou2023neuron, zhou2024evolutionary} target transfer granularity, fine-tuning only the most task-relevant neurons via importance ranking and evolutionary optimization. Complementary directions include margin-driven temporal triggers for model updates \cite{Lechowicz_2025_QoTEstimationMarginDriven} and feature-selection-based domain adaptation for lightweight classifiers \cite{aladin2025automated}. Despite their contributions, these approaches rely on sequential training pipelines and require explicit fine-tuning on target-domain data.

Beyond TL, researchers have explored architectural modifications to achieve generalization.
In \cite{Gao_2024_GeneralizationCognitiveOptical}, authors enhance generalization by decomposing networks into modular components that cascade to match lightpath structure, though requiring explicit topology knowledge at inference.
Similarly, \cite{Liu_2025_ModuleEnhanceGeneralization} introduces a lightweight adapter module for end-to-end systems, while \cite{Wang_2024_MultiSpanOpticalPower} inserts trainable loss models between pre-trained EDFA components. These structural methods require architectural re-mapping for each new configuration.

Representation learning through metric learning has also demonstrated potential for optical networks.
In \cite{Natalino_2019_OneshotLearningModulation}, a Siamese convolutional neural network (CNN) learns to compare constellation diagrams rather than classify them directly, achieving 100\% accuracy for both seen and unseen modulation formats.
This concept has been extended to fault management, where \cite{Gao_2024_FaultTracingBased} applies Siamese networks for root cause analysis, maintaining 83\% accuracy on imbalanced datasets where traditional artificial neural networks (ANNs) fail, while \cite{Natalino_2026_UnifiedSiameseLearning} proposed a multi-similarity Siamese framework that achieves 99\% accuracy for zero-day anomaly detection.
However, these techniques have primarily addressed classification problems, leaving generalizable representations unexplored.


Unlike prior optical-network TL work that relies on explicit topology knowledge or target-domain fine-tuning, our approach learns a task-aligned latent space directly from source-only supervision, making it deployable on networks for which no labeled operational data has yet been collected. A joint contrastive and classification training scheme simultaneously shapes representations to be compact within QoT classes and aligned with the downstream decision boundary, avoiding the sub-optimality of sequential pipelines and the need for extensive target-domain labels.

\begin{figure*}[t]
    \centering
    \includegraphics[width=0.99\linewidth]{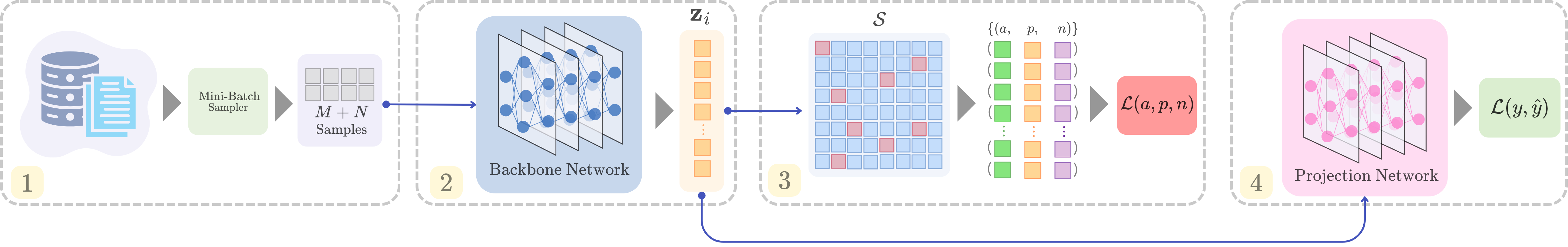}
    \caption{\footnotesize{Framework of representation learning for QoT lightpath. (1) Dataset is first sampled by a mini-batch sampler. (2) The backbone neural network computes a representation $\mathbf{z}_i$ for each sample in the batch. (3) The matrix $\mathcal{S}$ comprises pairwise similarity between all instances in the batch, which are used to mine informative triplets. The contrastive loss operates on these triplets to minimize an objective that maximizes the similarity between instances of the same class, and minimizes that of different classes. (4) A projection network learns to predict representations labels using a binary loss function}}
    \label{fig:architecture}
\end{figure*}

\section{Methodology}


\subsection{Problem Definition}
We address the problem of generalization of ML models under domain shift, i.e., mismatches between training and deployment data distributions. 
We define two datasets: a \emph{source} dataset, denoted by $\mathcal{D}_{src}$, and a \emph{target} dataset, denoted by $\mathcal{D}_{tgt}$. Our aim is to enhance the generalization capability of the ML model by learning task-relevant relationships that remain stable under domain shifts.

The source dataset $\mathcal{D}_{src}$ consists of labeled samples drawn from a source distribution $\mathcal{P}$ and is defined as:
\begin{equation}
\mathcal{D}_{src} = \left\{ \left(\mathbf{x}_i, y_i \right) \right\}_{i=1}^{N_{src}}, \quad (\mathbf{x}_i, y_i) \sim \mathcal{P}(\mathcal{X},\mathcal{Y})
\end{equation}
where $\mathbf{x}_i \in \mathcal{X}$ represents the feature vector describing a lightpath configuration and $y_i \in \mathcal{Y}$ denotes the corresponding label.

Similarly, the target dataset $\mathcal{D}_{tgt}$ is drawn from a different data distribution $\mathcal{Q}$, with $\mathcal{P} \neq \mathcal{Q}$, and is defined as:
\begin{equation}
\mathcal{D}_{tgt} = \left\{ \left(\mathbf{x}_j, y_j \right) \right\}_{j=1}^{N_{tgt}}, \quad (\mathbf{x}_{j}, y_{j}) \sim \mathcal{Q}.
\end{equation}

The discrepancy between $\mathcal{P}$ and $\mathcal{Q}$ captures domain shifts arising from differences in network topologies, devices, physical-layer conditions, or operational configurations. While the marginal distributions of the input features may differ across domains, the underlying learning task is assumed to remain unchanged. 

To this end, the goal is to learn a representation that captures task-relevant relationships shared across domains, while reducing sensitivity to domain-specific variations. We hypothesize that the learned model is capable of achieving improved generalization performance when deployed in previously unseen target domains due to focusing on invariant task semantics rather than domain-dependent correlations.

We define a model $M_\theta: \mathcal{X} \xrightarrow{f} \Phi \xrightarrow{g} \mathcal{Y}$, parameterized by $\theta=(\theta_1,\theta_2)$ representing parameters of learnable mapping functions $f$ and $g$ respectively, which projects inputs $\mathbf{x} \in \mathcal{X}$ into a representation space $\mathbf{z} \in \Phi$, that is used later to obtain its corresponding class. The objective is to minimize the expected prediction loss over unseen samples drawn from the target data distribution while being trained primarily on $\mathcal{D}_{src}$ and possibly a limited number of samples from $\mathcal{D}_{tgt}$.

\subsection{Proposed Architecture}
We propose an architecture based on a \emph{joint contrastive learning and classification}, as illustrated in Fig.~\ref{fig:architecture}. The architecture leverages deep metric learning (DML) \cite{hoffer2015deep} to learn a latent representation space in which samples corresponding to lightpaths are organized according to their QoT status (class).

The proposed architecture is organized into four tightly integrated functional blocks, denoted as \emph{Data Module}, \emph{Backbone Network and Latent Representation}, \emph{Mining and parameter Learning}, and \emph{Joint Projection and Classification}, numbered from 1 to 4 in Fig.~\ref{fig:architecture}, respectively, each addressing a specific aspect of the learning pipeline. In particular, the architecture is designed to \emph{(i)} mitigate class imbalance effects in the training data, \emph{(ii)} extract informative feature representations from raw lightpath descriptions, \emph{(iii)} learn a task-aligned latent manifold that captures meaningful relationships among samples, and \emph{(iv)} perform robust classification.

\textbf{Data Module.} Datasets pertaining to lightpath QoT estimation are imbalanced, often dominated by \lq \lq healthy" lightpaths (Class 1) with fewer instances of failure (Class 0). Standard random sampling can lead to model bias toward the majority class. To mitigate this, the Data Module employs a class-conditional sampler \cite{schroff2015facenet}. At each training iteration, the sampler constructs a mini-batch of size $B=M+N$, where $M$ and $N$ denote the number of samples drawn from the negative and positive classes, respectively. 

\textbf{Backbone Network and Latent Representation.} Inspired by recent advances in DML \cite{wang2019multi}, we employ a backbone network (BN) $f_{\theta_1}$, a deep neural network (DNN) parameterized by $\theta_1$. For an input sample $\mathbf{x}_i$ (representing lightpath features), the backbone maps the input to a $d$-dimensional latent vector $\mathbf{z}_i = f_{\theta_1}(\mathbf{x}_i) \in \Phi \subseteq \mathbb{R}^d$, i.e., a compact representation that encodes the essential characteristics of the lightpath relevant to the learning task.


Unlike standard classifiers, our backbone prioritizes representation learning, i.e., mapping inputs to a latent encoding $z_i$ that isolates the fundamental impairment factors determining QoT. This ensures the model captures task-relevant signal properties while filtering out redundant, domain-specific dependencies on network topology or configuration.

\textbf{Mining and Parameter Learning.} To ensure robustness across domains, our goal is to shape a latent geometry $\Phi$ that is invariant to network-specific variations. This is realized via a DML strategy that dynamically \emph{``mines''} triplets to cluster similar lightpaths and separate dissimilar ones \cite{hadsell2006dimensionality}. We compute a pairwise similarity matrix $\mathcal{S}$ online within each mini-batch to continuously optimize the embedding space; this enforces decision boundaries based on stable signal quality factors instead of transient domain artifacts.

The similarity $\mathcal{S}_{ij}$ between two instances $(\mathbf{z}_i,\mathbf{z}_j)$ is higher if the pair is positive (having same status) and lower if it is negative (having different status). $\mathcal{S}_{ij}$ is defined as: $
    \mathcal{S}_{ij} = \text{sim}(f_{\theta_{1}}(\mathbf{x}_i),f_{\theta_{1}}(\mathbf{x}_j)) = \text{sim}(\mathbf{z}_i,\mathbf{z}_j)$
where sim($\cdot$,$\cdot$) represents the cosine similarity.

In this work, we opt for online mining which, unlike offline mining \cite{hadsell2006dimensionality}, does not induce a lot of computation since it is performed on the fly on each batch at the end of the forward pass (as shown in Fig~\ref{fig:architecture}). Many online mining strategies have been proposed in the literature \cite{musgrave2020pytorch}. Recently, authors in \cite{wang2019multi} proposed \lq\lq \emph{Multi-Similarity Miner}'' (MS-Miner), a pair mining strategy that considers multiple pairwise similarities in the mining process. Specifically, MS-Miner constructs a triplet of samples from the batch $(a, p, n)$ consisting of an anchor $a$, a hard positive $p$ (a sample with the same label that appears unusually distinct from $a$), and a hard negative $n$ (an opposite-class sample that appears confusingly similar to $a$). By focusing on these difficult edge cases rather than easy samples, the model learns robust decision boundaries. The associated Multi-Similarity Loss (MS-Loss) \cite{wang2019multi} aggregates these pairwise relationships to optimize the latent structure, defined as:

\begin{equation}
\small
\begin{aligned}
\mathcal{L}_{CL}
&= \frac{1}{N} \sum_{i=1}^N \Bigg\{
\frac{1}{\alpha} \log\Bigg[1 + \sum_{j\in \mathcal{P}_i} e^{-\alpha(\mathcal{S}_{ij}-m)}\Bigg] \\
&\qquad\qquad
+ \frac{1}{\beta} \log\Bigg[1 + \sum_{k\in \mathcal{N}_i} e^{\beta(\mathcal{S}_{ik}-m)}\Bigg]
\Bigg\}
\label{eq_loss}
\end{aligned}
\end{equation}
where for each instance $\mathbf{z}_i$ in the batch, $\mathcal{P}_i$ represents the set of indices $\{j\}$ that form a positive pair with $\mathbf{z}_i$ and $\mathcal{N}_i$ the set of indices $\{k\}$ that form a negative pair with $\mathbf{z}_i$. $\alpha, \beta$ and $m$ represent constants that control the weighting scheme \cite{wang2019multi}.

\textbf{Joint Projection and Classification}
We introduce a \emph{Projection Network} (PN) $g_{\theta_{2}}$, a secondary neural network that maps the latent vector $z_i$ to a class probability $ \hat{y}_i = g_{\theta_{2}}(\mathbf{z}_i)$.

The PN is trained with a standard binary cross-entropy loss $\mathcal{L}_{CE}(y, \hat{y})$ using the ground-truth labels. In essence, this module serves as a classifier on top of the contrastively learned features. During training, it learns to interpret the embedding $\mathbf{z}_i$ and output the probability that the sample belongs to the positive class (e.g., acceptable QoT). Notably, this projection MLP is kept separate from the backbone in our design, allowing for both, a flexible training in isolation after the backbone is learned, or to train it jointly with the backbone. By employing this additional network, we also retain the option to fine-tune the high-level feature representation for the end task of QoT classification. 

\subsection{Training Strategy: Joint Optimization}
We now describe the training strategy. 
The literature commonly adopts a \emph{separate} training approach, where the BN is first trained with a contrastive objective and then frozen while a downstream classifier is learned. We observe that this strategy often leads to sub-optimal generalization, because freezing the backbone yields representations that are overly optimized for \emph{intra-class} compactness at the expense of the \emph{inter-class} separability needed by the classifier to establish robust decision boundaries.

To address this limitation, we propose a \emph{joint optimization} training technique, in which both the BN and the projection (or classification) networks are updated simultaneously during training. This technique allows gradients from both the contrastive and classification objectives to jointly shape the latent space, and hence, promote representations that are not only compact within classes but also well aligned with the downstream decision task, ultimately leading to improved robustness and generalization under domain shifts.

The total loss function of the joint training is defined as:
\begin{equation}
    \mathcal{L}_{total} = \gamma \mathcal{L}_{CL} + \eta \mathcal{L}_{CE}  
\end{equation}
where $\gamma$ and $\eta$ are hyperparameters.

\begin{table*}[t]
\centering
\caption{\footnotesize Cross-dataset evaluation across Sources D1, D2, and D3. Best result per column in \textbf{bold}; second-best \underline{underlined}.}
\label{tab:transfer_horizontal}
\scriptsize 
\setlength{\tabcolsep}{0pt} 
\begin{tabular*}{\textwidth}{@{\extracolsep{\fill}} l | cccc cccc | cccc cccc | cccc cccc}
\toprule
 & \multicolumn{8}{c|}{\textbf{Source: D1}} 
 & \multicolumn{8}{c|}{\textbf{Source: D2}} 
 & \multicolumn{8}{c}{\textbf{Source: D3}} \\
 
 & \multicolumn{4}{c}{Target: D2} & \multicolumn{4}{c|}{Target: D3} 
 & \multicolumn{4}{c}{Target: D1} & \multicolumn{4}{c|}{Target: D3} 
 & \multicolumn{4}{c}{Target: D1} & \multicolumn{4}{c}{Target: D2} \\
\cmidrule{2-5} \cmidrule{6-9} \cmidrule{10-13} \cmidrule{14-17} \cmidrule{18-21} \cmidrule{22-25}

Model & Acc & MF1 & PR & ROC & Acc & MF1 & PR & ROC 
      & Acc & MF1 & PR & ROC & Acc & MF1 & PR & ROC 
      & Acc & MF1 & PR & ROC & Acc & MF1 & PR & ROC \\
\midrule
RF   
& 0.593 & 0.515 & 0.552 & 0.594 & 0.567 & 0.468 & 0.630 & 0.623 
& 0.981 & 0.980 & 0.971 & 0.973 & 0.990 & 0.990 & 0.995 & 0.981 
& 0.589 & 0.515 & 0.549 & 0.672 & 0.739 & 0.722 & 0.658 & 0.772 \\

Extr 
& 0.575 & 0.487 & 0.542 & 0.578 & 0.558 & 0.452 & 0.601 & 0.584 
& 0.962 & 0.940 & 0.956 & 0.971 & \textbf{0.994} & \textbf{0.993} & 0.990 & \underline{0.993} 
& 0.586 & 0.517 & 0.547 & 0.657 & 0.578 & 0.487 & 0.542 & 0.549 \\

LR   
& 0.550 & 0.495 & 0.532 & 0.550 & 0.619 & 0.473 & \underline{0.715} & \underline{0.699} 
& 0.620 & 0.612 & 0.818 & 0.680 & 0.855 & 0.854 & 0.855 & 0.855 
& 0.552 & 0.522 & 0.528 & 0.590 & 0.676 & 0.666 & 0.611 & 0.684 \\

XGB  
& 0.602 & \underline{0.534} & 0.542 & 0.578 & \underline{0.628} & 0.487 & 0.683 & 0.676 
& \textbf{0.986} & \underline{0.982} & 0.972 & \textbf{0.985} & \underline{0.992} & \underline{0.992} & 0.986 & 0.992 
& \textbf{0.982} & \textbf{0.982} & \textbf{0.980} & \textbf{0.991} & \textbf{0.985} & \textbf{0.983} & \textbf{0.983} & \textbf{0.985} \\

CatB 
& \underline{0.604} & 0.532 & \underline{0.558} & \textbf{0.604} & 0.577 & \underline{0.485} & 0.700 & 0.577 
& \underline{0.985} & \textbf{0.985} & \textbf{0.984} & 0.975 & 0.991 & 0.991 & 0.984 & 0.898 
& 0.597 & 0.523 & 0.553 & 0.598 & 0.834 & 0.830 & 0.751 & 0.892 \\

MLP  
& 0.500 & 0.330 & 0.518 & 0.511 & 0.500 & 0.343 & 0.564 & 0.564 
& 0.583 & 0.508 & 0.749 & 0.745 & 0.748 & 0.737 & 0.889 & 0.877 
& 0.500 & 0.500 & 0.333 & 0.500 & 0.500 & 0.333 & 0.500 & 0.500 \\

Separate 
& 0.500 & 0.333 & 0.499 & 0.499 & 0.500 & 0.333 & 0.500 & 0.500 
& 0.892 & 0.892 & 0.935 & 0.933 & 0.824 & 0.854 & 0.894 & 0.894 
& 0.500 & 0.333 & 0.500 & 0.500 & 0.506 & 0.347 & 0.504 & 0.506 \\
\midrule
Ours 
& \textbf{0.674} & \textbf{0.644} & \textbf{0.656} & \underline{0.592} & \textbf{0.641} & \textbf{0.564} & \textbf{0.756} & \textbf{0.716} 
& 0.982 & 0.979 & \underline{0.973} & \underline{0.977} & \textbf{0.994} & 0.974 & \textbf{0.996} & \textbf{0.995} 
& \underline{0.672} & \underline{0.674} & \underline{0.759} & \underline{0.712} & \underline{0.841} & \underline{0.843} & \underline{0.898} & \underline{0.922} \\
\bottomrule
\end{tabular*}
\end{table*}

\section{Experimental Results}
\subsection{Experimental Settings}
We assess generalizability in scenarios where the target network provides no training data (zero-shot) and where small amounts of labelled data are incrementally made available. Our study uses three datasets ($D1$, $D2$, $D3$) collected from different optical network topologies described in \cite{bergk2022qotdataset} and available in \cite{fraunhofer_qot_collection}.

The three topologies differ in structural and operational properties: they span from 14-node regional meshes to 37-node national networks, with mean link lengths ranging from roughly 200\,km to 700\,km and average node degree between 2.7 and 4.4, producing markedly different distributions of per-lightpath span count, accumulated ASE noise, and nonlinear interference.
%
On top of this structural diversity,
the three datasets differ in two operational dimensions: transceiver mode and class balance. $D1$ uses Online Transceiver Mode, which permits flexible transceiver configurations and yields a higher failure rate (24.2\% negative class), whereas $D2$ and $D3$ use Predefined Transceiver Mode with much lower failure rates (7.3\% and 6.3\%, respectively). $D3$ also covers six data rates (50–-300\,Gb/s), against three (100/200/400\,Gb/s) for $D1$ and $D2$. These differences induce domain shifts rooted in topology, physical-layer impairments, configuration space, and class prior, providing a rigorous testbed for cross-domain robustness.
By training on one source $D_i$ and inferring on the others ($D_j, D_k$), we evaluate six cross-topology transfer settings: $D1 \to \{D2, D3\}$, $D2 \to \{D1, D3\}$, $D3 \to \{D1, D2\}$.

\paragraph*{Baselines:}
We compare against $(i)$ a \emph{Separate} training pipeline, in which the representation learning blocks are optimized first and then frozen before training the classifier, $(ii)$ a supervised Multi-Layer Perceptron (MLP) trained solely with cross-entropy, to isolate the contribution of the proposed contrastive component and joint optimization, and $(iii)$ competitive baselines for tabular learning, namely Random Forest (RF), ExtraTrees (Extr), Logistic Regression (LR), XGBoost (XGB), and CatBoost (CatB). 

\paragraph*{Evaluation metrics:}
For all experiments, we report Accuracy (Acc), Macro-F1 (MF1), Precision–Recall AUC (PR-AUC) and ROC-AUC.\footnote{In this imbalanced scenario, standard accuracy can be misleading; a trivial model could achieve 99\% accuracy simply by predicting \lq \lq Healthy" for every connection, failing to detect any faults. Therefore, we prioritize MF1, which treats both classes equally to ensure failures are not ignored, and PR-AUC, which specifically evaluates the \emph{reliability} of failure detection against false alarms. ROC-AUC is included for standard reference.}

\subsection{Cross-Dataset Generalization Results}
Table~\ref{tab:transfer_horizontal} reports the cross-dataset evaluation for our approach and the baselines. Our approach is robust across transfer directions. It matches the high performance of tree ensembles in favorable scenarios (e.g., Source D2) while delivering significant improvements in challenging cases where baselines suffer severe degradation (e.g., Source D1), mitigating the sharp drops observed under difficult domain shifts. As a result, our method achieves the highest stability, mitigating the sharp performance drops observed in standard ML models during difficult domain shifts.
Because the MS-Loss depends only on relative pairwise similarities, two lightpaths with different ASE accumulation or span counts but comparable SNR margin are pulled into the same cluster, so the representation is largely insensitive to dataset-specific amplifier gain profiles and link-length distributions.

\begin{figure*}
    \centering
    \setlength{\abovecaptionskip}{3pt}
    \includegraphics[width=1\linewidth]{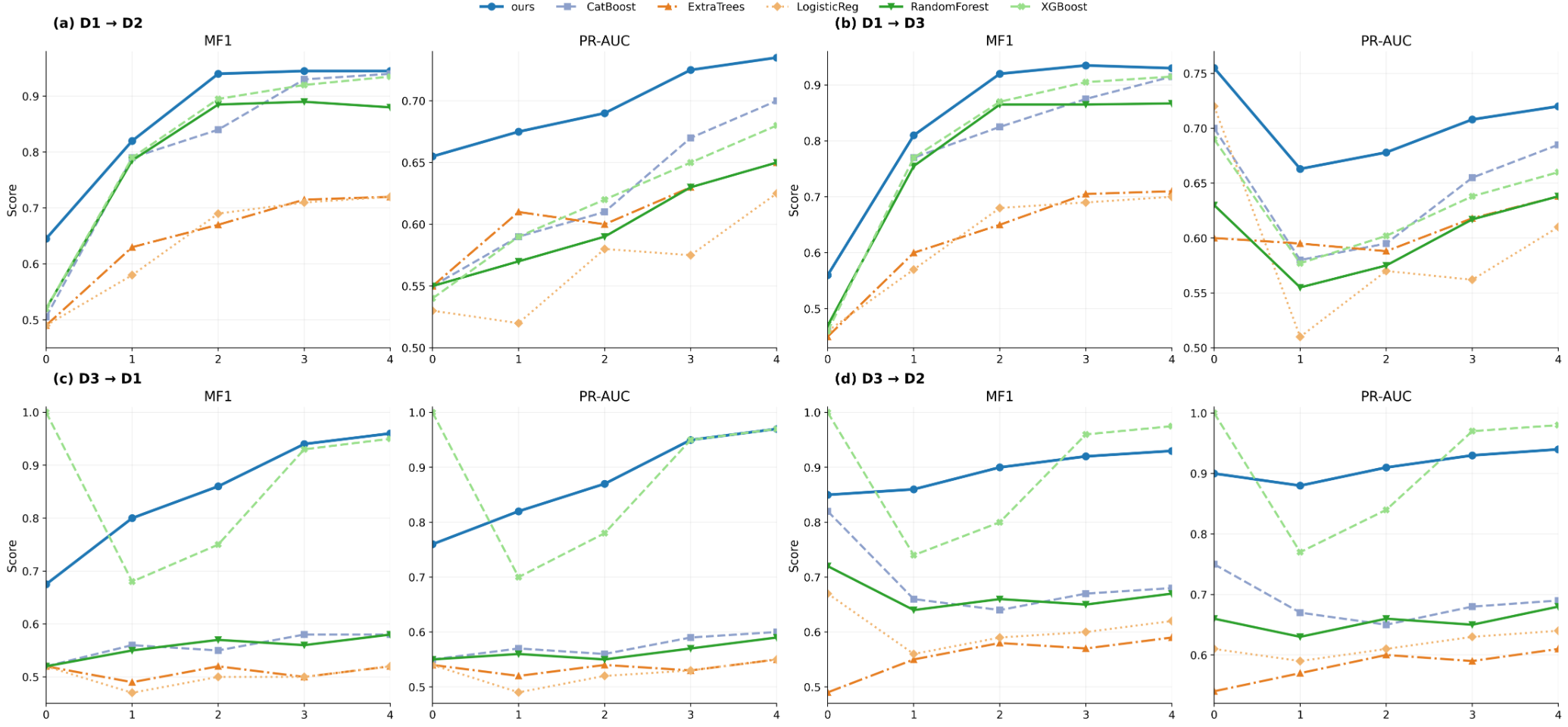}
    \caption{\footnotesize{Performance comparison in terms of MF1 and PR-AUC across (a) D1 $\rightarrow$ D2 (TL), (b) D1 $\rightarrow$ D3 (TR), (c) D3 $\rightarrow$ D1 (BL), and (d) D3 $\rightarrow$ D2 (BR).}}
    \label{fig:comparison_results}
\end{figure*}

We now examine each case in more detail. 

\paragraph*{Source $D1$:} 

$D1$'s Online Transceiver Mode and 24.2\% failure rate create a distributional profile unlike either Predefined-Mode target ($<$7.5\% failures), causing tree ensembles to overfit source-specific decision boundaries. Our approach improves MF1 from 0.534 to 0.644 and PR-AUC from 0.558 to 0.656 on $D1\!\rightarrow\!D2$, with similar gains on $D1\!\rightarrow\!D3$ (MF1: 0.487 $\to$ 0.564, PR-AUC: 0.715 $\to$ 0.756). These gains stem from the contrastive objective enforcing class-relative geometry that is invariant to the prior shift degrading baseline boundaries.

\paragraph*{Source $D2$:} 

$D2$ and $D3$ share Predefined Mode and near-identical class balance ($\sim7\%$); consequently, most tree-based baselines generalize strongly, while $D2\!\rightarrow\!D1$ transfers from a constrained to a richer configuration space. Our approach remains close to the top in both targets (MF1 $\geq$ 0.974), confirming that the contrastive objective does not sacrifice performance when domain shift is modest and that the learned representations remain discriminative.

\paragraph*{Source $D3$:} 
While $D3$ resembles $D2$ through its Predefined Mode and low failure rate (6.3\%), its broader data-rate diversity (six rates versus three) introduces a moderate distributional mismatch, particularly toward $D1$. In this setting, tree-based ensembles can achieve high performance; notably, XGBoost reaches an MF1 of 0.982 on $D3 \rightarrow D1$. However, this strength is not robust across source domains, as the same model degrades to an MF1 of 0.534 when transferred from D1, indicating source-specific and brittle decision boundaries. By contrast, our method trades peak performance for greater stability, yielding MF1 scores of 0.674 and 0.843 on $D3 \rightarrow D1$ and $D3 \rightarrow D2$, respectively, and maintaining a minimum PR-AUC of 0.656 across all six transfer directions, compared with 0.542 for XGBoost. This consistent performance floor is operationally significant, since real-world deployment is governed more by worst-case than by best-case generalization.

\paragraph*{Separate vs. Joint.:} Comparing our proposed joint training to \emph{Separate}, we see that the \emph{Separate} pipeline exhibits systematic failures under domain shift (e.g., MF1 $0.33$--$0.35$ with D3 as Source), supporting the motivation for joint optimization: freezing the representation can lead to embeddings that are insufficiently aligned with the downstream QoT decision boundary in unseen domains.

\subsection{Few-shot Target Adaptation via Data Injection}

To further assess deployability under domain shift, we study a \emph{target-data injection} setting where a small fraction of labeled samples from the target topology becomes available after model deployment. For each transfer direction $D_x\!\rightarrow\!D_y$, we consider injection ratios $i\in\{1,2,3,4\}\%$ of labeled target samples. 
This range reflects realistic early-deployment conditions in which only a handful of labeled lightpaths can be probed before the model is put in service; beyond $\sim5\%$ the regime approaches conventional supervised training on the target network and the cross-domain question becomes less meaningful.
A distinction holds in this experiment: for the baselines, there is no notion of ``fine-tuning''; hence, each operating point is obtained by \emph{retraining from scratch} on the union of the full source dataset $D_x$ and $i\%$ of the target dataset $D_y$. In contrast, our approach admits \emph{warm-start adaptation}: we first train on $D_x$ and then fine-tune the joint model using only the injected $i\%$ target samples. We report Macro-F1 (MF1) and PR-AUC as functions of the injected ratio (Fig.~\ref{fig:comparison_results}).

Across all reported transfers, our method exhibits a consistent and typically monotonic improvement as more target data are injected, indicating that the learned representation can be effectively \emph{refined} toward the target distribution with very limited supervision. Importantly, the gains are most pronounced at extremely small ratios. For $D1\!\rightarrow\!D2$, MF1 increases from $0.644$ (zero-shot) to approximately $0.82$ with only $1\%$ target data, and reaches $0.94$ at $2\%$; PR-AUC improves from $0.656$ to $0.68$ at $1\%$ and to $0.73$--$0.74$ at $4\%$. A similar pattern is observed for $D1\!\rightarrow\!D3$, where MF1 rises from $0.564$ to $0.81$ at $1\%$ and $0.92$ at $2\%$. While PR-AUC shows a transient decrease when moving from $0\%$ ($0.756$) to $1\%$ ($0.66$), it then steadily recovers with additional target samples, reaching $0.71$ by $3$--$4\%$. These results indicate that our model \emph{converges rapidly} toward high target performance, often requiring only $1$--$2\%$ labeled target data to close most of the cross-domain gap.\\
Another key observation is that several strong baselines behave \emph{non-monotonically} under small injections, i.e., adding a small amount of target data and retraining can temporarily degrade performance. This effect is visible when $D3$ is the source. For $D3\!\rightarrow\!D1$, XGB achieves near-ceiling zero-shot performance (MF1 $=0.982$, PR-AUC $=0.980$), yet drops when only $1\%$ target data are mixed and the model is retrained (MF1 $0.68$, PR-AUC $0.70$), before recovering at larger injections. In contrast, our model improves smoothly from MF1 $=0.674$ and PR-AUC $=0.759$ (zero-shot) to $0.80/0.82$ at $1\%$ and $0.86/0.87$ at $2\%$, reaching $0.94/0.96$ by $3\%$. The same phenomenon is observed for $D3\!\rightarrow\!D2$: XGB starts strong (MF1 $=0.983$, PR-AUC $=0.983$) but degrades markedly at $1\%$ (MF1 $0.74$, PR-AUC $0.77$) and remains below our method in the low-data regime. Our approach, instead, maintains stable improvements from MF1 $=0.843$ and PR-AUC $=0.898$ (zero-shot) to $0.86/0.88$ at $1\%$ and $0.90/0.91$ at $2\%$. These results highlight a practical advantage of the proposed pipeline: \emph{early-stage adaptation} (when only a limited number of labeled target samples can be collected) benefits from fine-tuning, whereas retraining-based baselines may suffer from instability.\\
Finally, we provide an overall comparison. In the low-injection regime ($1$--$2\%$), our approach is consistently the best or among the best across the shown transfers, typically outperforming all retrain-from-scratch baselines while using the same limited amount of target labels. At higher injections ($3$--$4\%$), some baselines, notably XGB in $D3\!\rightarrow\!D2$, eventually recover and can surpass our model after requiring substantially more target data (e.g., MF1/PR-AUC returning to $0.96$--$0.98$ at $3$--$4\%$). This behavior is consistent with the broader robustness claim of this work: while a specific baseline may excel in a particular transfer once sufficient target supervision is available, our joint contrastive pipeline delivers the most reliable and data-efficient adaptation, achieving strong target performance with minimal labeled target data and without the pronounced degradations observed for retraining-based methods at small injection ratios.

\section{Conclusion}
In this work, we propose a joint contrastive–classification learning approach to address the limited robustness of ML models under cross-domain deployment in optical networks. The method jointly optimizes representation learning and task performance to capture features that remain stable across heterogeneous domains. Experimental results on cross-dataset QoT estimation show that our approach matches strong baselines in favorable transfer scenarios while significantly improving robustness and worst-case performance under challenging domain shifts. Operationally, the resulting model can serve as a QoT pre-screening block fed by telemetry from the monitoring plane (per-span power, OSNR, amplifier gain), with retraining triggered only when the injected-sample PR-AUC drops below a deployment-defined threshold.


\bibliographystyle{IEEEtran}
\bibliography{references}

@IEEEtranBSTCTL{BSTcontrol,
  CTLuse_forced_etal       = "yes",
  CTLmax_names_forced_etal = "1",
  CTLnames_show_etal       = "1"
}

@article{Gao_2024_GeneralizationCognitiveOptical,
  author  = {Hanyu Gao and Xiaoliang Chen and Chao Lu and Zhaohui Li},
  title   = {On the generalization of cognitive optical networking applications using composable machine learning},
  journal = {Journal of Optical Communications and Networking},
  year    = {2024},
  volume  = {16},
  number  = {6},
  pages   = {631--643},
  doi     = {10.1364/JOCN.514981}
}

@inproceedings{Natalino_2019_OneshotLearningModulation,
  author       = {Carlos Natalino and Aleksejs Udalcovs and Lena Wosinska and Oskars Ozolins and Marija Furdek},
  title        = {One-Shot Learning for Modulation Format Identification in Evolving Optical Networks},
  booktitle    = {OSA Advanced Photonics Congress (AP)},
  year         = {2019},
  pages        = {JW4A.2},
}

@article{Liu_2025_ModuleEnhanceGeneralization,
  author  = {Zheng Liu and Tiegen Liu and Jian Zhao and Joshua Uduagbomen and Yulin Wang and Sergei Popov and Tianhua Xu},
  title   = {A Module to Enhance the Generalization Ability of End-to-End Deep Learning Systems in Optical Fiber Communications},
  journal = {Journal of Lightwave Technology},
  year    = {2025},
  volume  = {43},
  number  = {2},
  pages   = {596--601},
  doi     = {10.1109/JLT.2024.3466977}
}

@article{aladin2025automated,
  author    = {Sandra Aladin and Lena Wosinska and Christine Tremblay},
  title     = {Automated, Interpretable and Efficient {ML} Models for Real-World Lightpaths' Quality of Transmission Estimation},
  journal   = {IEEE Open Journal of the Communications Society},
  year      = {2025},
  volume    = {6},
  pages     = {9785--9801},
  doi       = {10.1109/OJCOMS.2025.3635533},
}

@article{zhou2024evolutionary,
  author  = {Yuhang Zhou and Zhiqun Gu and Jiawei Zhang and Yuefeng Ji},
  title   = {Evolutionary neuron-level transfer learning for {QoT} estimation in optical networks},
  journal = {Journal of Optical Communications and Networking},
  year    = {2024},
  volume  = {16},
  number  = {4},
  pages   = {432--448},
  doi     = {10.1364/JOCN.514618}
}

@article{usmani2024integrating,
  author  = {Fehmida Usmani and Ihtesham Khan and Arsalan Ahmad and Vittorio Curri},
  title   = {Integrating Knowledge Distillation and Transfer Learning for Enhanced {QoT}-Estimation in Optical Networks},
  journal = {IEEE Access},
  year    = {2024},
  volume  = {12},
  pages   = {156785--156802},
  doi     = {10.1109/ACCESS.2024.3485999}
}

@inproceedings{usmani2022transfer,
  author       = {Fehmida Usmani and Ihtesham Khan and Muhammad Umar Masood and Arsalan Ahmad and Muhammad Shahzad and Vittorio Curri},
  title        = {Transfer Learning Aided {QoT} Computation in Network Operating with the {400ZR} Standard},
  booktitle    = {International Conference on Optical Network Design and Modeling (ONDM)},
  year         = {2022},
  doi          = {10.23919/ONDM54585.2022.9782856}
}

@inproceedings{zhou2023neuron,
  author    = {Yuhang Zhou and Zhiqun Gu and Jiawei Zhang and Yuefeng Ji},
  title     = {Neuron-level Transfer Learning for {ANN}-based {QoT} Estimation in Optical Networks},
  booktitle = {Asia Communications and Photonics Conference / International Photonics and Optoelectronics Meetings (ACP/POEM)},
  year      = {2023},
  doi       = {10.1109/ACP/POEM59049.2023.10368618}
}

@inproceedings{Wang_2024_MultiSpanOpticalPower,
  author       = {Zehao Wang and Yue-Kai Huang and Shaobo Han and Ting Wang and Dan Kilper and Tingjun Chen},
  title        = {Multi-Span Optical Power Spectrum Prediction Using {ML}-based {EDFA} Models and Cascaded Learning},
  booktitle    = {Optical Fiber Communication Conference (OFC)},
  year         = {2024},
  pages        = {M1H.6},
  doi          = {10.1364/OFC.2024.M1H.6},
}

@inproceedings{Sadighi_2025_GeneralizabilityMLBasedClassification,
  author    = {Leyla Sadighi and Carlos Natalino and Stefan Karlsson and Marco Ruffini and Eoin Kenny and Lena Wosinska and Marija Furdek},
  title     = {Generalizability of {ML}-Based Classification of State of Polarization Signatures Across Different Bands and Links},
  booktitle = {European Conference on Optical Communications (ECOC)},
  year      = {2025},
  pages     = {Th.02.01.2},
  location  = {Copenhagen, Denmark},
  isbn      = {979-8-3315-9531-9},
  doi       = {10.1109/ECOC66593.2025.11263096}
}

@inproceedings{Akbari_2025_LeveragingSharedData,
  author    = {Hassan Akbari and Xiao Ma and Behnam Shariati and Pooyan Safari and Angela Mitrovska and Johannes K. Fischer and Stephan Pachnicke and Jasper M{\"u}ller and Ronald Freund},
  title     = {Leveraging Shared Data and Models for {ML}-Based {QoT} Estimation: Toward Standardized and Generalizable Models},
  booktitle = {European Conference on Optical Communications (ECOC)},
  year      = {2025},
  pages     = {W.04.01.3},
  doi       = {10.1109/ECOC66593.2025.11263378}
}

@inproceedings{Natalino_2026_UnifiedSiameseLearning,
  author    = {Carlos Natalino and Fl{\'a}via Pessoa Monteiro and Paolo Monti},
  title     = {A Unified {Siamese} Learning Framework for Zero-Day Anomaly Detection and Classification in Optical Networks},
  booktitle = {Optical Fiber Communication Conference (OFC)},
  year      = {2026},
  note      = {to appear},
  location  = {Los Angeles, CA, USA},
  url       = {https://research.chalmers.se/en/publication/549889}
}

@inproceedings{Gao_2024_FaultTracingBased,
  author    = {Yuxuan Gao and Bingli Guo and Yu Zhou and Yuting Ma and Kuan Yan and Shanguo Huang},
  title     = {Fault Tracing Based on {Siamese} Neural Network for Optical Networks},
  booktitle = {IEEE Opto-Electronics and Communications Conference (OECC)},
  year      = {2024},
  doi       = {10.1109/OECC54135.2024.10975324}
}

@inproceedings{Lechowicz_2025_QoTEstimationMarginDriven,
  author       = {Piotr Lechowicz and Carlos Natalino and Paolo Monti},
  title        = {{QoT} Estimation with Margin-Driven Transfer Learning in Time-Varying Optical Networks},
  booktitle    = {Optical Fiber Communication Conference (OFC)},
  year         = {2025},
  pages        = {M1J.5},
  doi          = {10.1364/OFC.2025.M1J.5},
}

@inproceedings{hoffer2015deep,
  author       = {Elad Hoffer and Nir Ailon},
  title        = {Deep Metric Learning Using Triplet Network},
  booktitle    = {International Workshop on Similarity-Based Pattern Recognition},
  year         = {2015},
  pages        = {84--92},
}

@inproceedings{wang2019multi,
  author    = {Xun Wang and Xintong Han and Weilin Huang and Dengke Dong and Matthew R. Scott},
  title     = {Multi-Similarity Loss with General Pair Weighting for Deep Metric Learning},
  booktitle = {Proceedings of the IEEE/CVF Conference on Computer Vision and Pattern Recognition},
  year      = {2019},
  pages     = {5022--5030}
}

@inproceedings{schroff2015facenet,
  author    = {Florian Schroff and Dmitry Kalenichenko and James Philbin},
  title     = {FaceNet: A Unified Embedding for Face Recognition and Clustering},
  booktitle = {Proceedings of the IEEE Conference on Computer Vision and Pattern Recognition},
  year      = {2015},
  pages     = {815--823}
}

@inproceedings{hadsell2006dimensionality,
  author    = {Raia Hadsell and Sumit Chopra and Yann LeCun},
  title     = {Dimensionality Reduction by Learning an Invariant Mapping},
  booktitle = {IEEE Computer Society Conference on Computer Vision and Pattern Recognition (CVPR'06)},
  year      = {2006},
  volume    = {2},
  pages     = {1735--1742},
}

@article{musgrave2020pytorch,
  author  = {Kevin Musgrave and Serge Belongie and Ser-Nam Lim},
  title   = {{PyTorch} Metric Learning},
  journal = {arXiv preprint arXiv:2008.09164},
  year    = {2020}
}

@article{bergk2022qotdataset,
  author  = {Geronimo Bergk and Behnam Shariati and Pooyan Safari and Johannes K. Fischer},
  title   = {{ML}-assisted {QoT} estimation: a dataset collection and data visualization for dataset quality evaluation},
  journal = {Journal of Optical Communications and Networking},
  volume  = {14},
  number  = {3},
  pages   = {43--55},
  year    = {2022},
  doi     = {10.1364/JOCN.442733}
}

@misc{fraunhofer_qot_collection,
  author       = {{Fraunhofer Heinrich-Hertz-Institut}},
  title        = {{QoT} Dataset Collection},
  year         = {2026},
  howpublished = {\url{https://www.hhi.fraunhofer.de/en/pn-software/qot-dataset-collection.html}},
  note         = {Accessed: 2026-03-29}
}

\end{document}